# Modelling and Control of Combustion Phasing in Dual-Fuel Compression Ignition Engines

Wenbo Sui[1], Jorge Pulpeiro González[1] and Carrie M. Hall[1]*

1: Illinois Institute of Technology

## Abstract:

Dual fuel engines can achieve high efficiencies and low emissions but also can encounter high cylinder-to-cylinder variations on multi-cylinder engines. In order to avoid these variations, they require a more complex method for combustion phasing control such as model-based control. Since the combustion process in these engines is complex, typical models of the system are complex as well and there is a need for simpler, computationally efficient, control-oriented models of the dual fuel combustion process. In this paper, a mean-value combustion phasing model is designed and calibrated and two control strategies are proposed. Combustion phasing is predicted using a knock integral model, burn duration model and a Wiebe function and this model is used in both an adaptive closed loop controller and an open loop controller. These two control methodologies are tested and compared in simulations. Both control strategies are able to reach steady state in 5 cycles after a transient and have steady state errors in CA50 that are less than ±0.1 crank angle degree (CAD) with the adaptive control strategy and less than ±1.5 CAD with the model-based feedforward control method.

## Introduction

In the transportation area, regulations continue to require lower emissions while the market simultaneously demands higher fuel efficiencies. These requirements have led to many new technological developments including variable geometry turbochargers, variable valve actuation systems and advanced exhaust aftertreatment systems. Dual-fuel capabilities have also been explored and been shown to have the potential to increase efficiencies while producing much lower emissions of nitrogen oxides and particulate matter than conventional diesel engines [1]. Dual fuel combustion strategies typically use two fuels with different reactivities to create in-cylinder reactivity stratification that provide greater control of combustion phasing and pressure rise rate [2]. These benefits of dual fuel engines have been explored on single cylinder [3] and multi-cylinder engines [4, 5, 6, 7, 8]. As with conventional internal combustion engines, optimal combustion phasing is critical to maintain high efficiencies on dual fuel engines, and as such, control of this parameter is paramount [9, 10].

Due to the increasingly complex configuration of these engines, there is interest in using model-based control strategies to accurately control the engine processes including combustion phasing. Combustion phasing on conventional diesel engines has been captured using a variety of models. In [11, 12, 13], a model based on the Shell auto-ignition method was used to make precise prediction of the start of combustion (SOC), but this model requires a detailed CFD analysis for calibration. The knock integral model (KIM), which was first proposed by Livengood and Wu to predict the knock of a spark ignition engine [14], has also been utilized to predict SOC [15,16,17] on diesel engines. In [14], Hillion et al. employed a KIM to predict SOC and the end of the cold flame. Other factors such as emissions including CO and NO mass flow rates have also been estimated using this model [15,16]. In addition, the KIM has been modified by Shahbakhti to include the impact of varying air-fuel ratio and EGR fraction on SOC in a homogenous charge compression ignition (HCCI) engine [18]. Different forms of the Arrhenius factor in the knock integral were also studied in [19] in order to predict and control the SOC and location of peak pressure in a HCCI engine.

Efforts to expand these models to dual fuel compression ignition engines include extensions of semi-empirical models such as the Wiebe function [20] and probability density function (PDF) based models. In [21] and [22], a PDF based stochastic reactor model is leveraged to model the combustion process of a dual-fuel HCCI engine and is experimentally verified. This model characterizes the combustion process including chemical kinetics and inhomogeneous compositions and temperatures. With this PDF based model, Bhave et al. are able to estimate SOC. In [23], a single zone combustion model with detailed chemical mechanism was successfully used to obtain an accurate estimation of SOC for dual fuel engines. However, this model does not properly estimate burn duration and has a high computation time. In [20], the ignition delay of a dual fuel engine is captured, but other aspects of combustion phasing are not addressed. In contrast, this work seeks to model the combustion phasing of a dual fuel engine including SOC and CA50.

Combustion phasing control has been accomplished using traditional PID (Proportion Integral Derivative) controllers as well as underlying models and feedback from additional sensors. In [24], a modified KIM model is simplified via an empirical correlation and an observer-based state-feedback controller is implemented to control SOC. In [25], Olsson et al. proposed a closed-loop PID controller to achieve the desired CA50 with short response time in a dual-fuel HCCI engine. The PID gains were tuned based on a sensitivity function that was developed based on the experimental data in a HCCI dual-fuel engine. A similar method is investigated by Maurya et al. in [26]. A simple PID controller was utilized to control CA50 with pressure rise rate limitation based on fuel blend ratio. Strandh et al. leveraged a PID controller and a LQG (linear Quadratic Gaussian) controller to control CA50 based on ion measurements and in-cylinder pressure feedback [27]. Another PID controller based on discrete linearized model was posted by Ott et al. in [28]. The controller adjusted the injection time and duration of diesel fuel for controlling the combustion phasing and the maximum in-cylinder pressure. In reference [29], three PI (Proportion Integral) controllers based on a map-based RCCI model were used to control the CA50 and the IMEP (indicated mean effective pressure). Kondipati et al. have utilized a PI controller to track the desired CA50 by controlling the fuel mixture ratio and injection timing [30]. In order to reject the disturbance, a look-up table feedforward controller was used along with the PI feedback controller.

Unlike these prior studies, this work explores a simplification to a modified KIM model that makes it more suitable for real-time control. In this work SOC and CA50 are modelled and two control approaches are developed. Since combustion efficiency is a strong function of CA50, control of these parameters are critical [31]. The model predicts the SOC using a modified knock integral model (MKIM) and computes the burn duration (BD) using a separate burn duration model. CA50 is later calculated using these two parameters. The integral model is simplified into a non-linear model that is calibrated using simulation data.

An adaptive feedback controller is designed to control the CA50 based on this non-linear model using a measurement of CA50. Simulation results show that the combustion phasing can reach steady state in 10 cycles after a transient and the steady state error in CA50 is less than ±0.1 CAD. Although there are some commercial diesel engines that have in-cylinder sensors that could estimate CA50 values, other production engines do not. Thus, use of this control strategy may be more limited. To overcome this disadvantage, an additional control strategy based on open-loop feedforward model-based control was developed. This control technique can give an appropriate start of injection (SOI) based on the simple non-linear model. Simulation results for this second control strategy showed that CA50 settled to the steady state value in 10 cycles when the operating conditions were changed and the steady state errors were less than ±1.5 CAD.

The paper describes the combustion phasing prediction model. After the experimental and simulation setup are described, the model development and calibration procedure is discussed as well as the development of the two different control strategies. Next, the results of the simulations are presented and analyzed. Finally, some conclusions are made regarding the performance of the control techniques used in this work.

# Dual-Fuel Engine Experimental and Simulation Model Setup

Fig.1 shows the configuration of the dual-fuel engine studied in this work. In this engine, ambient air is compressed (by a compressor) and mixed with recirculated exhaust gasses and a low reactivity fuel in the intake manifold. After the intake stroke, a high reactivity fuel is injected into the cylinder and combustion occurs. After the exhaust stroke, the burned gases go to the exhaust manifold, where portion goes to the exhaust gas recirculation (EGR) flow path towards the intake manifold and the rest passes through the turbine, generating the power necessary to drive the compressor. In this work, natural gas was used as the low reactivity fuel and diesel was the high reactivity fuel.

To calibrate the CA50 prediction model and evaluate its accuracy, it is compared to experimental results as well as simulation results from GT-ISE (Gamma Technology's Integrated Simulation Environment). The simulation model used in this work has been previously used in [8] and [9] and is described briefly here. The engine specifications are shown in Table 1.

This simulation model was developed and validated based on 20 experiments at different operating points. It uses both a direct-injection combustion model and traditional port-injection combustion model to capture the dual fuel combustion event. Results from 10 diesel only experiments were used to properly calibrate direct-injection portion in the GT-ISE simulation model, and 10 dual fuel experiments with diesel and natural gas were used to calibrate the port-injection portion [9]. This simulation model has a ±1.8 CAD uncertainty compared with experimental results and is used here to simulate the combustion process of the dual-fuel engine over a wide range of operating points as shown in Table 2. A total of 1054 simulations at different operating conditions were tested, and the coefficients in the CA50 model are calibrated using these simulations.

# Combustion Phasing Model

A block diagram of the combustion phasing prediction model for dual-fuel engines is shown in Fig. 2. This model contains three parts: a SOC model, burn duration model and CA50 model. First, the SOC is predicted by a modified knock integral model based on the SOI (start of diesel injection), pressure and temperature at IVC ($P_{IVC}$ and $T_{IVC}$), engine speed ($N$), and EGR fraction, as well as the equivalence ratios of diesel ($\phi_{DI}$) and natural gas ($\phi_{NG}$). Meanwhile, the burn duration (BD) model is estimated based on the EGR fraction, residual gas fraction ($X_r$), the diesel and natural gas equivalence ratios. Afterwards, the CA50 is predicted based on SOC and BD by a CA50 model developed from Wiebe function.

## SOC Prediction Model

A KIM captures the relationship between IVC and SOC as

$$\int_{IVC}^{SOC} \frac{\tau}{N} d\theta = 1 \tag{1}$$

where $IVC$ denotes intake valve closing, $SOC$ indicates the start of combustion, $N$ represents the engine speed, and $\tau$ denotes the Arrhenius function.

The Arrhenius function can take different forms but is often a function of fuel equivalence ratios, compression temperature and pressure. In this work, the Arrhenius function is given by

$$\tau = \frac{1}{(a_1 EGR + a_2)\phi^{a_3}} \exp\left(-\frac{a_4 P^{a_5}}{T}\right) \tag{2}$$

in which $EGR$ represents the EGR fraction, $\phi$ is the fuel equivalence ratio, $T$ and $P$ are the dynamic temperature and pressure, and $a_1, a_2, a_3, a_4$ and $a_5$ are the empirical constants. In this engine, natural gas is port-injected and diesel is direct-injected. As such, the equivalence ratio term in Eqn. (2) is not constant in dual-fuel engines. Since diesel is injected at SOI, the equivalence ratio term should have different forms from IVC to SOI and SOI to SOC. Therefore, the Arrhenius function between IVC and SOI can be written as in Eqn. (3), and the Arrhenius function between SOI and SOC is captured by Eqn. (4).

$$\tau_1 = \frac{1}{(b_1 EGR + b_2)\phi_{NG}^{b_3}} \exp\left(-\frac{b_4 P^{b_5}}{T}\right) \tag{3}$$

$$\tau_2 = \frac{1}{(c_1 EGR + c_2)(\phi_{NG}^{c_3} + \phi_{DI}^{c_4})} \exp\left(-\frac{c_5 P^{c_6}}{T}\right) \tag{4}$$

Here $\phi_{NG}$ is the equivalence ratio of natural gas and $\phi_{DI}$ indicates the equivalence ratio of diesel, as defined in Eqns. (5) and (6), respectively.

$$\phi_{NG} = \left(\frac{m_{NG}}{m_{air}}\right) / \left(\frac{m_{NG}}{m_{air}}\right)_{st} \tag{5}$$

$$\phi_{DI} = \left(\frac{m_{diesel}}{m_{air}}\right) / \left(\frac{m_{diesel}}{m_{air}}\right)_{st} \tag{6}$$

In these equations, $m_{NG}$ and $m_{diesel}$ are the natural gas injection mass and diesel injection mass each cycle, and $m_{air}$ is the mass of air entering in the engine cylinder. The subscript $st$ indicates stoichiometric conditions.

Substituting Eqns. (3) and (4) into Eqn. (1), the KIM can be rewritten as

$$\int_{IVC}^{SOI} \frac{\exp\left(-\frac{b_4 P^{b_5}}{T}\right)}{(b_1 EGR + b_2) N \phi_{NG}^{b_3}} d\theta + \int_{SOI}^{SOC} \frac{\exp\left(-\frac{c_5 P^{c_6}}{T}\right)}{(c_1 EGR + c_2) N (\phi_{NG}^{c_3} + \phi_{DI}^{c_4})} d\theta = 1 \quad (7)$$

As demonstrated in [24] and confirmed in this study, the first integral term in Eqn. (7) is much smaller than the second integral term in Eqn. (7), because the activation energy of natural gas $b_4$ is much higher than the activation energy of the mixture of the fuels $c_5$ [24, 32, 33]. As such, the exponential term in the second integral is almost $10^9$ times as the exponential term in the first integral and the first term can be ignored in the modified knock integral model of dual-fuel engines. Even though the period from IVC to SOI is larger than the period from SOI to SOC, the first term in Eqn. (7) is neglectable. Therefore, the knock integral model of dual-fuel engine SOC can be described by

$$\int_{SOI}^{SOC} \frac{\exp\left(-\frac{c_5 P^{c_6}}{T}\right)}{(c_1 EGR + c_2) N (\phi_{NG}^{c_3} + \phi_{DI}^{c_4})} d\theta = 1 \quad (8)$$

It should be noted that this knock integral model assumes that combustion does not occur from IVC to SOC. During this process, wall heat losses are also ignored from IVC to SOC as is typical in knock integral models [34].

## Burn Duration Prediction Model
Burn duration is captured by

$$BD = c_7 (1 + X_d)^{c_8} (\phi_{NG}^{c_9} + \phi_{DI}^{c_{10}}) \quad (9)$$

where $X_d$ represents the dilution fraction, $\phi_{NG}$ and $\phi_{DI}$ are the equivalence ratios of natural gas and diesel, and $c_7$, $c_8$ $c_9$ and $c_{10}$ are constant parameters. The dilution fraction $X_d$ will include both recirculated exhaust gas and residuals. As such, $X_d$ can be represented as

$$X_d = EGR + X_r \quad (10)$$

in which $EGR$ denotes the EGR fraction, and $X_r$ indicates the residual fraction. The residual fraction $X_r$ is defined by

$$X_r = \frac{m_r}{m_{air} + m_{NG} + m_{diesel} + m_{EGR}} \quad (11)$$

where $m_r$ represents the mass of residual air in the cylinder from the last cycle, $m_{air}$ is the mass of air entering the cylinder, $m_{EGR}$ denotes the mass of recirculated exhaust gas, and $m_{NG}$ and $m_{diesel}$ are mass of natural gas and diesel, respectively. With values or estimates of $EGR$, $X_r$ and the equivalence ratios of fuels, the burn duration can be estimated.

## CA50 Prediction Model
Based on the prediction of SOC and BD, a Wiebe function can be employed to predict fraction of fuel burned as

$$x_b(\theta) = 1 - \exp\left(-a \left[\frac{\theta - SOC}{BD}\right]^b\right) \quad (12)$$

where $\theta$ is the crank angle, $x_b$ indicates the mass fraction of burned fuel, $SOC$ represents the timing of SOC in crank angle, $BD$ denotes the burn duration in crank angle, and $a$ and $b$ are constant coefficients.

CA50 can be calculated by evaluating Eqn. (12) when $x_b$ equals 0.5. Thus, Eqn. (12) can be written as

$$0.5 = 1 - \exp\left(-a\left[\frac{CA50 - SOC}{BD}\right]^b\right) \tag{13}$$

Rearranging, CA50 can be predicted by

$$CA50 = SOC + \left[\frac{\ln 2}{a}\right]^{1/b} BD. \tag{14}$$

Substituting Eqn. (9) into Eqn. (14), CA50 can be captured by

$$CA50 = SOC + c_{11}(1 + X_d)^{c_8}\left(\phi_{NG}^{c_9} + \phi_{DI}^{c_{10}}\right) \tag{15}$$

in which $c_{11}$ is defined as

$$c_{11} = \left[\frac{\ln 2}{a}\right]^{1/b} c_7. \tag{16}$$

Using these submodels, SOC can be estimated by Eqn. (8), and CA50 can be predicted by Eqn. (15).

## CA50 Model Simplification

Although SOC and CA50 can be predicted by Eqns. (8) and (15), the integral model is still too complex for application in the typical control design. Therefore, it is necessary to simplify the integral model in Eq. (8) for further control design work. The model can be simplified by assuming that in one specific combustion cycle, equivalence ratios, engine speed as well as EGR fraction can be treated as constants. With that assumption, the equivalence ratio, engine speed and EGR fraction terms can be taken out from the integral term as reflected in Eqn. (17).

$$\frac{1}{(c_1 EGR + c_2)N\left(\phi_{NG}^{c_3} + \phi_{DI}^{c_4}\right)} \int_{SOI}^{SOC} \exp\left(-\frac{c_5 P^{c_6}}{T}\right) d\theta = 1 \tag{17}$$

The dynamic pressures and temperatures in Eqn. (17) can be captured by a polytropic correlation, stemming from the pressure and temperature at IVC.

Typically, to achieve the optimal CA50 in dual-fuel engines, the SOI should occur around -20° to 5° aTDC, and the SOC occurs from 1 – 10 CAD after SOI. With such small crank angle changes from SOI to SOC, the cylinder volume does not change dramatically during this period. Therefore, the dynamic pressures and temperatures from SOI to SOC can be treated as constant values during this period, and they can be replaced by the pressure and temperature at SOI. While this simplification will result in a slight loss in accuracy, it allows the integral equation to be simplified as

$$\frac{1}{(c_1 EGR + c_2)N\left(\phi_{NG}^{c_3} + \phi_{DI}^{c_4}\right)} \exp\left(-\frac{c_5 P_{SOI}^{c_6}}{T_{SOI}}\right)(SOC - SOI) = 1 \tag{18}$$

where $P_{\text{SOI}}$ and $T_{\text{SOI}}$ are the pressure and temperature at SOI and they can be derived by

$$P_{\text{SOI}} = P_{\text{IVC}} \left(\frac{V_{\text{IVC}}}{V_{\text{SOI}}}\right)^{k_c} \tag{19}$$

and

$$T_{\text{SOI}} = T_{\text{IVC}} \left(\frac{V_{\text{IVC}}}{V_{\text{SOI}}}\right)^{k_c-1}. \tag{20}$$

SOC can be predicted by rearranging Eqn. (18) as

$$SOC = SOI + (c_1 EGR + c_2) N \left(\phi_{\text{NG}}^{c_3} + \phi_{\text{DI}}^{c_4}\right) \exp\left(\frac{c_5 P_{\text{SOI}}^{c_6}}{T_{\text{SOI}}}\right). \tag{21}$$

Substituting Eqn. (21) into Eqn. (15), CA50 can be captured by

$$\begin{aligned} CA50 = {} & SOI + (c_1 EGR + c_2) N \left(\phi_{\text{NG}}^{c_3} + \phi_{\text{DI}}^{c_4}\right) \exp\left(\frac{c_5 P_{\text{SOI}}^{c_6}}{T_{\text{SOI}}}\right) \\ & + c_{11}(1 + X_d)^{c_8} \left(\phi_{\text{NG}}^{c_9} + \phi_{\text{DI}}^{c_{10}}\right). \end{aligned} \tag{22}$$

Thus, the CA50 in dual-fuel engines can be predicted by Eqn. (22) with the SOI, EGR fraction, engine speed, equivalence ratios of fuels, pressure and temperature at SOI and dilution fraction.

# CA50 Model Validation

## Calibration of CA50 Prediction Model

GT-ISE simulations were utilized to calibrate the coefficients in the CA50 prediction model. The time step of these simulations is 0.1 CAD. The root mean square error (RMSE) of CA50 was minimized by a batch gradient descent algorithm for the CA50 prediction model based on all 1054 simulations. These simulations were tested at steady state. The model calibration procedure is shown in Fig. 3. First, an initial guess of the parameters provided the initial conditions for the calibration and the gradient of RMSE was set to zero. For each simulation, the CA50 was predicted using the CA50 model. After that, the RMSE and gradient of RMSE of this simulation were calculated. The RMSE and RMSE gradient of all simulations was updated using the gradient of RMSE of the current simulation. After update, the parameters were revised based on the RMSE gradient. The calibration iteration was stopped when the RMSE could no longer be decreased.

The calibrated parameters of the CA50 prediction model in Eqn. (22) are given in Table 3.

## Validation of CA50 Prediction Model

Based on calibrated parameters in Table 3, the SOC and CA50 can be predicted by Eqns. (21) and (22). The predicted SOCs and CA50s are compared with the simulation results in Fig. 4. In Fig. 4 (a), the x axis is the SOC from the GT-ISE simulation, and the y axis is the predicted SOC from Eqn. (21). The dashed blue lines are ±1 CAD error limits. The standard deviation of the prediction error is 0.2999 CAD, and the maximum error is 1.4852 CAD.

Similarly, Fig. 4 (b) compares the predicted CA50 to the simulation CA50 and the dashed blue lines are ±1 CAD error limits. The standard deviation of the CA50 prediction error is 0.5317 CAD, and the maximum error is 2.1902 CAD. Comparison with Fig. 4 shows that the error of CA50 prediction is not only from errors in the prediction of SOC, but also from inaccuracies in the prediction of BD. The errors in the SOC prediction model are mainly due to the fact that heat transfer was ignored from IVC to SOC.

While these errors could be reduced with more complex models, the results from Fig. 4 demonstrate that the Eqns. (21) and (22) can predict SOC and CA50 with the error less than 1.5 CAD and 2.2 CAD, respectively. This is a sufficient level of accuracy to be useful for dual-fuel engine combustion phasing control applications for the operating region parameterized and errors in this range should only have a minor impact on the thermal efficiency of the engine. Based on this CA50 prediction model, an adaptive feedback control strategy and a model-based open-loop control strategy were investigated. Note that this engine can experience cylinder-to-cylinder variations in CA50 particularly when late IVC timings are used. More information on CA50 variations in this dual fuel engine can be found in [36]. When high CA50 variations occur, more sophisticated control approaches may be required, but consideration of cylinder-to-cylinder variations was outside the scope of this study.

## Adaptive Feedback Control Strategy

An adaptive control strategy was expected to provide precise CA50 control. However, this strategy requires some method of CA50 feedback. While this type of feedback is not common, it was investigated in this work to demonstrate the usefulness of such an approach and provide a good reference with which to compare the simpler open loop method that will be discussed later.

## State-Space Model Derivation

To design an adaptive controller dual-fuel engine combustion phasing control, a state-space model is used and is derived from the CA50 dynamic model in Eqn. (22). In this work, CA50 is the control reference $y$, and the SOI is the control input $u$. Thus, the state-space model can be captured by

$$y = u + \alpha x_1 + \beta x_2 \tag{23}$$

in which

$$\alpha = (c_1 EGR + c_2) \exp\left(\frac{c_5 P_{SOI}{}^{c_6}}{T_{SOI}}\right) \tag{24}$$

$$\beta = c_{11}(1 + X_d)^{c_8} \tag{25}$$

$$x_1 = N\big(\phi_{NG}^{c_3} + \phi_{DI}^{c_4}\big) \tag{26}$$

and

$$x_2 = \big(\phi_{NG}^{c_9} + \phi_{DI}^{c_{10}}\big). \tag{27}$$

In the CA50 state-space model shown in Eqns. (23) – (27), the states $x_1$ and $x_2$ change from cycle to cycle and can be computed from the engine speed and the equivalence ratios. The parameters $\alpha$ and $\beta$, which are influenced by $P_{\text{SOI}}$, $T_{\text{SOI}}$ and dilution fraction $X_d$, must be updated from cycle to cycle.

## Adaptive Controller Design

Based on the CA50 state-space model, an adaptive controller was designed and a block diagram of the control strategy is shown in Fig. 5.

As shown in Fig. 5, the optimal CA50 serves as the reference input for the controller. Measurements or estimates of the engine speed and equivalence ratio of each fuel are used to calculate the states with Eqns. (26) and (27). Based on the reference CA50 and the states, the adaptive controller gives the appropriate SOI to engine control unit (ECU) for the diesel injection. The actual CA50 is sent back to the adaptive controller as the feedback to update the parameters $\alpha$ and $\beta$ in Eqns. (24) and (25).

Using the CA50 prediction model in Eqn. (23), the appropriate SOI ($u$) can be found based on the parameters $\alpha$ and $\beta$, states $x_1$ and $x_2$ and the reference output $y_d$ (the desired CA50). Rewriting Eqn. (23), the input $u$ can be given as

$$u = y_d - \alpha x_1 - \beta x_2. \tag{28}$$

Since the parameters $\alpha$ and $\beta$ vary from cycle to cycle and include unmeasured parameters ($P_{\text{SOI}}$, $T_{\text{SOI}}$ and dilution fraction $X_d$), the controller can only calculate $u$ based on estimated values rather than the actual values of $\alpha$ and $\beta$. Replacing the actual values of $\alpha$ and $\beta$ by the observed values $\bar{\alpha}$ and $\bar{\beta}$, Eqn. (28) can be written as

$$u = y_d - \bar{\alpha} x_1 - \bar{\beta} x_2. \tag{29}$$

To ensure accurate observed values of parameters $\alpha$ and $\beta$, an observer is designed based on gradient descend algorithm. The error function is defined as the root mean squared error (RMSE) of the CA50, and it is given by

$$E = \frac{1}{2}(y - \bar{y})^2 \tag{30}$$

where $E$ expresses the RMSE, $y$ is the actual output and $\bar{y}$ represents the observed output. The observed output $\bar{y}$ can be captured by

$$\bar{y} = u + \bar{\alpha} x_1 + \bar{\beta} x_2. \tag{31}$$

Comparison of Eqn. (31) and Eqn. (29) shows that if the input $u$ and the estimated parameters $\bar{\alpha}$ and $\bar{\beta}$ are constants in steady state, the estimated output $\bar{y}$ should be the same as the desired output $y_d$. Thus, Eqn. (30) can be rewritten by replacing $\bar{y}$ by $y_d$ as

$$E = \frac{1}{2}(y - y_d)^2. \tag{32}$$

With the error function in Eqn. (32), the partial derivatives of the observed parameters are

$$\frac{\partial E}{\partial \bar{\alpha}} = -x_1(y - y_d) \tag{33}$$

$$\frac{\partial E}{\partial \bar{\beta}} = -x_2(y - y_d). \tag{34}$$

With the gradient descent algorithm, the observed parameters $\bar{\alpha}$ and $\bar{\beta}$ can be updated from cycle to cycle by following equations

$$\bar{\alpha}(k+1) = \bar{\alpha}(k) - \eta \frac{\partial E}{\partial \bar{\alpha}} \tag{35}$$

$$\bar{\beta}(k+1) = \bar{\beta}(k) - \eta \frac{\partial E}{\partial \bar{\beta}} \tag{36}$$

where $\bar{\alpha}(k+1)$, $\bar{\beta}(k+1)$, $\bar{\alpha}(k)$ and $\bar{\beta}(k)$ are the observed parameters at the $k+1$ cycle and $k$ cycle, respectively; $\eta$ indicates the learning rate of the algorithm; and $\frac{\partial E}{\partial \bar{\alpha}}$ and $\frac{\partial E}{\partial \bar{\beta}}$ are the gradients given in Eqns. (33) and (34).

Substituting (33) into (35) and substituting (34) into (36), the parameter update equations are
$$\bar{\alpha}(k+1) = \bar{\alpha}(k) + \eta x_1 (y - y_d) \tag{37}$$
$$\bar{\beta}(k+1) = \bar{\beta}(k) + \eta x_2 (y - y_d). \tag{38}$$

In this work, the learning rate $\eta$ is
$$\eta = \frac{1}{x_1^2 + x_2^2}. \tag{39}$$

The learning rate $\eta$ in Eqn. (39) is chosen to achieve system stability, short settling time and small overshoot.

Substituting Eqn. (39) into Eqns. (37) and (38), the update equations for the observed parameters are

$$\bar{\alpha}(k+1) = \bar{\alpha}(k) + \frac{x_1}{x_1^2 + x_2^2} (y - y_d) \tag{40}$$

$$\bar{\beta}(k+1) = \bar{\beta}(k) + \frac{x_2}{x_1^2 + x_2^2} (y - y_d). \tag{41}$$

Eqns. (29), (40) and (41) summarize the adaptive feedback control system that can be used to track the desired CA50.

## Adaptive Control System Stability
The stability of the adaptive control system should be guaranteed before practical application in dual-fuel engines. A Lyapunov direct method is utilized to prove the stability of the strategy. Since the goal of the control system is to minimize the error between the actual CA50 and desired CA50, the Lyapunov function is chosen as
$$V[x(k)] = (y_d - y)^2. \tag{42}$$

For this chosen Lyapunov function, it can be shown that
$$\begin{cases} V[x(k)] = 0, \text{if } y_d - y = 0, \\ V[x(k)] > 0, \forall\, y_d - y \neq 0 \text{ and} \\ V[x(k)] \to \infty, \text{if } y_d - y \to \infty. \end{cases} \tag{43}$$

Substituting Eqns. (23) and (29) into Eqn. (42), the Lyapunov function can be captured by:

$$V[x(k)] = \left[x_1\big(\alpha(k) - \bar{\alpha}(k)\big) + x_2\big(\beta(k) - \bar{\beta}(k)\big)\right]^2 \tag{44}$$

To simplify Eqn. (44), the observer errors are described as

$$\tilde{\alpha}(k) = \alpha(k) - \bar{\alpha}(k) \tag{45}$$

$$\tilde{\beta}(k) = \beta(k) - \bar{\beta}(k) \tag{46}$$

Substituting Eqns. (45) and (46) into Eqn. (44), the Lyapunov function can be rewritten as

$$V[x(k)] = \left[x_1\tilde{\alpha}(k) + x_2\tilde{\beta}(k)\right]^2. \tag{47}$$

Similar to Eqn. (47), Lyapunov function at the $k+1$ cycle can be captured by

$$V[x(k+1)] = \left[x_1\tilde{\alpha}(k+1) + x_2\tilde{\beta}(k+1)\right]^2 \tag{48}$$

where

$$\tilde{\alpha}(k+1) = \alpha(k+1) - \bar{\alpha}(k+1) \tag{49}$$

$$\tilde{\beta}(k+1) = \beta(k+1) - \bar{\beta}(k+1). \tag{50}$$

Substituting Eqns. (40) and (41) into Eqns. (49) and (50) respectively, the errors of parameters observation at $k+1$ cycle can be given by Eqns. (51) - (52).

$$\tilde{\alpha}(k+1) = \alpha(k+1) - \bar{\alpha}(k) - \frac{x_1}{x_1^2 + x_2^2}(y - y_d) \tag{51}$$

$$\tilde{\beta}(k+1) = \beta(k+1) - \bar{\beta}(k) - \frac{x_2}{x_1^2 + x_2^2}(y - y_d) \tag{52}$$

Subtracting Eqn. (29) from Eqn. (23) yields

$$y - \bar{y} = x_1\big(\alpha(k) - \bar{\alpha}(k)\big) + x_2\big(\beta(k) - \bar{\beta}(k)\big). \tag{53}$$

Because the observed output $\bar{y}$ and desired $y_d$ are the same in steady state, $\bar{y}$ can be replaced by $y_d$ in Eqn. (53) giving

$$y - y_d = x_1\big(\alpha(k) - \bar{\alpha}(k)\big) + x_2\big(\beta(k) - \bar{\beta}(k)\big). \tag{54}$$

Substituting Eqns. (45) and (46) into Eqn. (54), it can be written as

$$y - y_d = x_1\tilde{\alpha}(k) + x_2\tilde{\beta}(k). \tag{55}$$

The observation errors at the $k+1$ cycle can be captured by substituting Eqn. (55) into Eqns. (51) and (52) resulting in

$$\tilde{\alpha}(k+1) = \alpha(k+1) - \bar{\alpha}(k) - \frac{x_1}{x_1^2 + x_2^2}\left[x_1\tilde{\alpha}(k) + x_2\tilde{\beta}(k)\right] \tag{56}$$

$$\tilde{\beta}(k+1) = \beta(k+1) - \bar{\beta}(k) - \frac{x_2}{x_1^2 + x_2^2}\left[x_1\tilde{\alpha}(k) + x_2\tilde{\beta}(k)\right]. \tag{57}$$

Since the parameters $\alpha$ and $\beta$ are constant during steady state,

$$\alpha(k+1) = \alpha(k) \tag{58}$$

and

$$\beta(k+1) = \beta(k). \tag{59}$$

Substituting Eqns. (58) and (59) into Eqns. (56) and (57) respectively, the parameter observation errors are

$$\tilde{\alpha}(k+1) = \alpha(k) - \bar{\alpha}(k) - \frac{x_1}{x_1^2 + x_2^2}[x_1\tilde{\alpha}(k) + x_2\tilde{\beta}(k)] \quad (60)$$

$$\tilde{\beta}(k+1) = \beta(k) - \bar{\beta}(k) - \frac{x_2}{x_1^2 + x_2^2}[x_1\tilde{\alpha}(k) + x_2\tilde{\beta}(k)]. \quad (61)$$

Substituting Eqns. (45) and (46) into Eqns. (60) and (61), these equations can be simplified to

$$\tilde{\alpha}(k+1) = \tilde{\alpha}(k) - \frac{x_1}{x_1^2 + x_2^2}[x_1\tilde{\alpha}(k) + x_2\tilde{\beta}(k)] \quad (62)$$

$$\tilde{\beta}(k+1) = \tilde{\beta}(k) - \frac{x_2}{x_1^2 + x_2^2}[x_1\tilde{\alpha}(k) + x_2\tilde{\beta}(k)]. \quad (63)$$

Further substituting Eqns. (62) and (63) into Eqn. (48), the Lyapunov function at the $k+1$ cycle is given by Eqn. (64).

$$V[x(k+1)] = \begin{bmatrix} x_1\left[\tilde{\alpha}(k) - \frac{x_1}{x_1^2 + x_2^2}[x_1\tilde{\alpha}(k) + x_2\tilde{\beta}(k)]\right] + \\ x_2\left[\tilde{\beta}(k) - \frac{x_2}{x_1^2 + x_2^2}[x_1\tilde{\alpha}(k) + x_2\tilde{\beta}(k)]\right] \end{bmatrix}^2 \quad (64)$$

Simplifying Eqn. (64) at steady state, it can be found that

$$V[x(k+1)] = 0. \quad (65)$$

Therefore, the difference between the Lyapunov function at the $k+1$ cycle and the $k$ cycle is captured in Eqn. (66).

$$V[x(k+1)] - V[x(k)] = -(y_d - y)^2 \quad (66)$$

According to Eqn. (66), the difference between the Lyapunov function at the $k+1$ cycle and $k$ cycle is negative if $y_d - y$ is not zero. Summarizing Eqn. (66) and Eqn. (43), it can be concluded that

$$\begin{cases} V[x(k)] = 0, \text{if } y_d - y = 0, \\ V[x(k)] > 0, \forall\, y_d - y \neq 0, \\ V[x(k)] \to \infty, \text{if } y_d - y \to \infty \text{ and} \\ V[x(k+1)] - V[x(k)] < 0, \forall\, y_d - y \neq 0. \end{cases} \quad (67)$$

According to the Lyapunov direct method [35], the output of combustion phasing control system is globally asymptotically stable.

## Feedforward Model-based Controller Design

Because in-cylinder pressure sensors, which can provide CA50 estimates, are not common on production engines, the adaptive feedback control strategy may be challenging. For dual-fuel engines without CA50 measurement, a model-based open-loop controller could still be applied. Such a controller is designed here to control combustion phasing by leveraging the dual fuel CA50 model. The block diagram of such a control system is shown in Fig. 6.

The reference CA50 is still the reference input. The appropriate SOI is calculated by a model-based open-loop controller with the reference input and the engine parameters including engine speed, equivalence ratios, EGR fraction, pressure and temperature at IVC. Using the CA50 prediction model shown in Eqn. (26), the SOI can be determined by

$$SOI = CA50_{ref} - (c_1 EGR + c_2)N(\phi_{NG}^{c_3} + \phi_{DI}^{c_4}) \exp\left(\frac{c_5 P_{SOI}^{c_6}}{T_{SOI}}\right) \\ - c_{11}(1 + X_d)^{c_8}(\phi_{NG}^{c_9} + \phi_{DI}^{c_{10}}) \tag{68}$$

in which $CA50_{ref}$ is the reference CA50. Engine speed ($N$) is typically available on production ECUs, and the equivalence ratios of fuels $\phi_{NG}$ and $\phi_{DI}$ can be calculated by Eqns. (5) and (6). EGR fraction ($EGR$) can be estimated based on the oxygen fraction at the intake and exhaust manifold as

$$EGR = \frac{x_{O2,amb} - x_{O2,exh}}{x_{O2,amb} - x_{O2,int}} \times 100\% \tag{69}$$

where $x_{O2,amb}$, $x_{O2,exh}$, $x_{O2,int}$ are the oxygen fractions of ambient air, at the intake manifold and at the exhaust manifold. If oxygen sensors are not available in the intake and exhaust manifolds, EGR flow can be estimated using methods such as those presented in [37].

Besides these measurable parameters, the dilution fraction ($X_d$) and the pressure and temperature at SOI ($P_{SOI}$ and $T_{SOI}$) cannot be measured directly from the sensors. Eqns. (19) and (20) can be used to derive $P_{SOI}$ and $T_{SOI}$ from the pressure and temperature at IVC ($P_{IVC}$ and $T_{IVC}$). This IVC pressure and temperature can be estimated based on measured intake manifold conditions or a direct measurement can be used. In this work, measured values of $P_{IVC}$ and $T_{IVC}$ are leveraged. Substituting Eqns. (19) and (20) into Eqn. (68), the SOI can be captured by

$$SOI = CA50_{ref} - (c_1 EGR + c_2)N(\phi_{NG}^{c_3} + \phi_{DI}^{c_4}) \exp\left[\frac{c_5\left[P_{IVC}\left(\frac{V_{IVC}}{V_{SOI}}\right)^{k_c}\right]^{c_6}}{T_{IVC}\left(\frac{V_{IVC}}{V_{SOI}}\right)^{k_c-1}}\right] \\ - c_{11}(1 + X_d)^{c_8}(\phi_{NG}^{c_9} + \phi_{DI}^{c_{10}}) \tag{70}$$

where $V_{SOI}$ is the volume of cylinder at SOI. Since the cylinder volume at a particular crank angle is known, $V_{SOI}$ can be found for a particular choice of SOI. However, if $V_{SOI}$ is left in Eqn. (70), the control law would need to complete several iterations to compute the appropriate SOI. To decrease the computation procedure of the control strategy, the $V_{SOI}$ from the previous cycle is used.

Meanwhile, the dilution fraction $X_d$ can be computed based on EGR fraction and residual gas fraction $X_r$ in Eqn. (10). While a variety of methods exist for estimating EGR fraction, the residual gas fraction $X_r$ cannot be measured directly and is difficult to estimate. The value of $X_r$ ranges from 0.02 to 0.05 for the conditions considered in this work and has much less variation than the EGR fraction. As such, it is reasonable to use the average value of $\bar{X}_r$ ($\bar{X}_r = 0.0329$) to calculate SOI. Note that if conditions such as positive valve overlap were considered, $X_r$ could be much larger and may need to be separately modeled. Thus, the model-based open-loop control law is given by

$$SOI = CA50_{\text{ref}} - (c_1 EGR + c_2)N(\phi_{\text{NG}}^{c_3} + \phi_{\text{DI}}^{c_4}) \exp\left[\frac{c_5\left[P_{\text{IVC}}\left(\frac{V_{\text{IVC}}}{V_{\text{SOI}}}\right)^{k_c}\right]^{c_6}}{T_{\text{IVC}}\left(\frac{V_{\text{IVC}}}{V_{\text{SOI}}}\right)^{k_c-1}}\right] \quad (71)$$

$$-c_{11}(1 + EGR + \bar{X}_r)^{c_8}(\phi_{\text{NG}}^{c_9} + \phi_{\text{DI}}^{c_{10}})$$

The model-based open-loop controller will use Eq. (71) to generate an appropriate SOI in an effort to track the desired CA50.

## Simulation and Analysis

In this section, the two controllers based on adaptive feedback control and model predictive control are evaluated in simulations. Each test will consider changes from one steady state point to another and six different cases will be considered. In these simulations, the precision of SOI is set as 0.1 CAD. The detailed settings for all six Cases are listed in the Appendix, while the simulation results are plotted in Fig. 7.

In the first case, the controllers' performance is evaluated during a change in reference CA50. Figure 7(a) shows the performance of both control algorithms during this change. In Fig. 7 (a), the red curve is the reference CA50, and the green and blue lines are the actual CA50 achieved with adaptive control and open-loop control, respectively. Fuel was not injected into the system in first 2 cycles so the CA50s in those 2 cycles are zero. In the first 5 seconds, both control systems reach their steady states in 5 cycles. The adaptive controller has a 0.8 CAD overshoot, and the steady state error varies from -0.075 to 0.035 CAD during steady state. This oscillation of the error is mainly from the assumed precision limitations of the injection rather than the control algorithm. In contrast, the open-loop controller has a -0.42 CAD steady state error without overshoot. The open-loop controller does not have such overshoot because it does not have a feedback iteration.

When the reference CA50 is changed, the actual CA50 is changed a cycle later because of the reference CA50 sampling method. At 5 seconds, the reference CA50 is still 8 CAD, and it has changed into 10 CAD at 5.001 seconds. Because the controller has already received the reference CA50 and calculated the SOI before the new reference CA50 is generated, the actual CA50 has a one cycle delay. After the transient, the CA50 from the adaptive control case reaches steady state in 2 cycles with a 0.27 CAD overshoot, and the steady state error is from -0.1 to 0.04 CAD. The open-loop controller does not have overshoot, and the steady state error is -0.30 CAD.

Next, the capability of the controllers to maintain the desired CA50 during an engine speed change was explored. The simulation result is plotted in Fig. 7 (b). As mentioned before, the fuel does not go into the system in the first 2 cycles, which leads to zero CA50 in the first 2 cycles. In the first 5 seconds, both control systems can reach their steady states in 5 cycles. The overshoot in the adaptive control system is 0.801 CAD. The steady state error mainly from the SOI precision is -0.073 – 0.030 CAD. Different from adaptive controller, the steady state in the open-loop control system is -0.420 CAD. After 5 seconds, the engine speed increases smoothly from 1200 RPM to 1500 RPM in 0.5 second. Both control strategies have reached their steady state in 2 cycles after the transient. The steady state errors are -0.033 – 0.077 CAD and -0.912 CAD in adaptive control system and open-loop control system, respectively. The higher steady state error in open-loop control system at the second operating condition is due to higher prediction error in the model.

In Case 3, the influence of a change in the natural gas equivalence ratio is evaluated by running simulation. Since the influence from natural gas equivalence ratio is similar with the effect by diesel equivalence ratio, only the equivalence ratio of natural gas is shown here. During the first 5 seconds, the error is fairly low with only a steady state error of -0.031 to 0.071 CAD with adaptive control system and -0.33 CAD steady state error with open-loop control. During the transient, the natural gas equivalence ratio moves from 0.3 to 0.5 in 0.5 second. The increasing equivalence ratio means that more natural gas has been mixed with the intake air. During the transient, the residual gas fraction and the EGR composition change due to the change of natural gas quantity. These changes also lead to changes in the intake gas properties, including the pressure and temperature at IVC. Despite this, good performance is seen with both controllers. After the transient, the CA50s from both systems take about 5 cycles to reach the new steady states but have errors of -0.103 to 0.004 CAD with adaptive control and -0.328 CAD with the open-loop strategy.

As operating conditions change, EGR fraction may also be expected to change quite dramatically and the impact of this is explored in Case 4. The simulation results are given in Fig. 7 (d), respectively. Steady state errors prior to the transient at 5 seconds are low at -0.074 – 0.040 CAD for adaptive control and -0.167 CAD for open-loop control. The EGR fraction changes from 0 to 0.5 during a 0.5 second ramp beginning at the 5 second mark and this leads to significant errors initially with both controllers. Since the air handling system typically is not capable of abrupt changes, this ramp is closer to the actual performance of the engine, but yet is still a very aggressive change. As shown in Eqns. (23) - (25), the EGR fraction is treated as a disturbance term in the adaptive control system. Thus, it takes about 10 cycles for the adaptive controller adjust to the ramp change in EGR fraction. After a 1.332 CAD overshoot from 5.7 to 5.8 seconds, the actual CA50 backs to a steady state around 8 CAD in 10 cycles. Unlike the adaptive controller, the CA50 with open-loop control decreases from 5 to 5.5 second with its lowest value 6.25 CAD, and then jumps to a high of 9.20 CAD at 5.9 seconds. This significant variation is related to the delayed effect of the EGR fraction change. The intake gas properties as well as the quantity of fuel injected change as a result of the EGR fraction change. However, these changes are several cycles later than the changes in EGR fraction. Because the EGR fraction values are directly used in the model leveraged by the open-loop control strategy, the delay causes more error during the transient. After the transient, the steady state error is again low and ranges from -0.058 CAD to 0.045 CAD with adaptive control system and is 0.875 CAD with open-loop control.

As in Case 5, a combined change including a ramp in natural gas equivalence ratio and engine speed is tested. The results are captured in Figure 6 (e). Similar to Case 2 and Case 3, CA50 in the adaptive control system achieves its steady state in 5 cycles with an overshoot 0.603 CAD. The steady state error is from -0.033 to 0.072 CAD. Meanwhile, the open-loop control system has an error -0.332 CAD. From 5 to 5.5 seconds, both engine speed and natural gas equivalence ratio are ramped to their second value. In the transient, the engine speed increases from 1200 RPM to 1500 RPM, and the equivalence ratio of natural gas goes from 0.3 to 0.5. These ramp changes cause a slightly oscillation for the adaptive control system. The CA50 in the open-loop system moved from 7.669 to 7.307 CAD during the transient with the maximum error is -0.836 CAD. Both controllers can reach steady state conditions at the second operating point before the 6 second mark. The steady states error of adaptive controller and open-loop controller are -0.063 to 0.046 CAD and 0.693 CAD, respectively.

A second combined change was considered for Case 6. In Case 6, EGR fraction changes along with engine speed and natural gas equivalence ratio. This case is essentially a combination of Case 4 and Case 5 and as such, the error also appears to be primarily an addition of the error in Case 4 and Case 5. In the first operating condition before 5 second, the steady state error of adaptive control system is from -0.026 to 0.079 CAD, while this error is 0.242 CAD for the open-loop control system. From 5 to 5.5 second, the engine speed, natural gas equivalence ratio and the EGR fraction were changed smoothly in ramp. The performance during this transient is similar to that of Case 4. The CA50 achieved with the adaptive controller reaches its peak value of 9.521 CAD at 5.7 second and goes back to steady state after. In the open-loop control system, the CA50 drops to a minimum of 6.08 CAD at 5.4 second and jumps to a maximum value of 9.90 CAD at 5.8 seconds. As explained in Case 4, delays in the air handling system cause higher model errors. After the transient, the steady state error in adaptive control system is -0.049 to 0.032, while the open-loop controller has a steady state error is 1.394 CAD. Comparison with Case 4 shows that the main error in Case 6 is likely from the change in EGR fraction.

These simulations show that both controllers can reach steady state quickly with a maximum steady state error less than ±0.1 CAD for adaptive control, and ±1.5 CAD for open-loop control. Still, both controllers cannot track the desired CA50 as well when the parameters of the engine change abruptly. This is especially significant in Case 4 and Case 6, which has a large EGR fraction change in only 0.5 second. However, such a big, sudden change does not typically happen in normal engine operation and Case 6 simply shows a worst-case scenario. As expected, the adaptive controller is more accurate that the open loop one, since it has a feedback signal, but since CA50 measurements are not available in most commercial diesel engines, the open-loop controller could be more widely used in stock diesel engines. Besides these operating conditions changes, the measurements error is another source which would highly influence on the controller performance.

## Discussion of Measurement Errors

Because the performance of the controllers is highly dependent on sensor measurements, measurement errors could significantly impact performance. The influence of the sensor measurement errors on the performance of controllers is studied in this section. First, the error response of the open-loop controller is investigated. Meanwhile, the adaptive controller is tested with CA50 measurements error in the simulation. After these tests, the impact of the anticipated errors in CA50 on the combustion efficiency is also discussed.

For such an open-loop model-based controller, measurement uncertainty including measurements errors, noise and cyclic variations will highly impact the control errors. To investigate this, the CA50 prediction error from $P_{IVC}$, $T_{IVC}$, EGR fraction, fuel equivalence ratios and $X_r$ have been studied. A semi-empirical model of $P_{IVC}$ and $T_{IVC}$ is presented in [47], and the uncertainties of $P_{IVC}$ and $T_{IVC}$ prediction error are ±0.036 bar and ±3.8 K, respectively. Therefore, $P_{IVC}$ is tested with ±0.05 bar error, and $T_{IVC}$ is tested with ±5 K error. The 1054 simulations which were used to calibrate and validate the combustion phasing prediction model were used to test the accuracy of the prediction model with measurements errors and the results are listed in Table 6. This sensitivity analysis only considers changes in one variable at a time and as such, cross effects by multiple-variables are not discussed. As shown in Table 6, measurement errors can produce up to 2.7 CAD error in CA50 prediction for the open-loop controller application even with fairly sizable errors in EGR and equivalence ratio.

The error response of the adaptive controller also needs to be analyzed. To evaluate the impact of sensor noise and measurement error in the adaptive control system, a simulation was run with CA50 measurement error. Lab grade pressure measurements revealed a 0.26 CAD uncertainty in CA50 and a 0.5 CAD uncertainty in CA50 measurement was used to test the error response of the adaptive controller. The settings of this test are the same as that for the first operating condition in Case 1. The results of the simulation are shown in Fig.8. Both the actual CA50 and the measured CA50 with noise are shown in the figure. The standard deviation of the actual CA50 error is 0.8512 CAD and the maximum error is 2.1766 CAD during the 10 second test.

In the future work, method for decreasing the impact of measurement noise should be considered. The addition of a filter or the change of the learning rate in Eqn. (49) may improve the error response in adaptive control system.

# Conclusion

In this paper, MIKM, BD model and a Wiebe function are used to predict the combustion phasing of a dual fuel engine utilizing diesel and natural gas. These models are combined and simplified into a non-linear model that is used to estimate CA50. This model was calibrated using data from simulations and two control strategies were developed based on this model. The adaptive feedback control uses the actual CA50 measured from in-cylinder sensors and has a maximum steady state error of less than ±0.1 CAD at different operating conditions. The open-loop controller has a maximum steady state error less than ±1.5 CAD for the same operating conditions as the adaptive feedback control. Thus, both control strategies have an acceptable performance. Also, the impact of measurement error and sensor noise on both control techniques was evaluated and both showed adequate performance. Since the knock integral model and Wiebe function can be widely used in different engines, the control strategies developed in this paper could be applied to a variety of CI engines. In future work, these strategies will be tested experimentally and the performance of the controllers during drive cycles will be also evaluated. The prediction model may also be able to be improved in future work by including a model of heat losses or more accurate models of the gas exchange processes or by leveraging a higher fidelity simulation model for calibration.

# Acknowledgement

This material is based upon work supported by the National Science Foundation under Grant No. 1553823.

# Appendix

Table A. Settings for Simulation Cases

| Case Number | Quantity | First Operating Condition | Second Operating Condition |
|---|---|---|---|
| 1 | Engine Speed (RPM) | 1200 | 1200 |
| | Average Temperature at Intake Manifold (K) | 300 | 300 |
| | Average Pressure at Intake Manifold (bar) | 2 | 2 |
| | Diesel Equivalence Ratio (-) | 0.4 | 0.4 |
| | Natural Gas Equivalence Ratio (-) | 0.4 | 0.4 |
|   | EGR fraction (-) | 0.25 | 0.25 |
|   | Reference CA50 (CAD) | 8 | 10 |
| 2 | Engine Speed (RPM) | 1200 | 1500 |
|   | Average Temp. at Intake Manifold (K) | 300 | 300 |
|   | Average Pressure at Intake Manifold (bar) | 2 | 2 |
|   | Diesel Equivalence Ratio (-) | 0.4 | 0.4 |
|   | Natural Gas Equivalence Ratio (-) | 0.4 | 0.4 |
|   | EGR fraction (-) | 0.25 | 0.25 |
|   | Reference CA50 (CAD) | 8 | 8 |
| 3 | Engine Speed (RPM) | 1200 | 1200 |
|   | Average Temperature at Intake Manifold (K) | 300 | 300 |
|   | Average Pressure at Intake Manifold (bar) | 2 | 2 |
|   | Diesel Equivalence Ratio (-) | 0.4 | 0.4 |
|   | Natural Gas Equivalence Ratio (-) | 0.3 | 0.5 |
|   | EGR fraction (-) | 0.25 | 0.25 |
|   | Reference CA50 (CAD) | 8 | 8 |
| 4 | Engine Speed (RPM) | 1200 | 1200 |
|   | Average Temperature at Intake Manifold (K) | 300 | 300 |
|   | Average Pressure at Intake Manifold (bar) | 2 | 2 |
|   | Diesel Equivalence Ratio (-) | 0.4 | 0.4 |
|   | Natural Gas Equivalence Ratio (-) | 0.4 | 0.4 |
|   | EGR fraction (-) | 0.0 | 0.5 |
|   | Reference CA50 (CAD) | 8 | 8 |
| 5 | Engine Speed (RPM) | 1200 | 1500 |
|   | Average Temperature at Intake Manifold (K) | 300 | 300 |
|   | Average Pressure at Intake Manifold (bar) | 2 | 2 |
|   | Diesel Equivalence Ratio (-) | 0.4 | 0.4 |
|   | Natural Gas Equivalence Ratio (-) | 0.3 | 0.5 |
|   | EGR fraction (-) | 0.25 | 0.25 |
|   | Reference CA50 (CAD) | 8 | 8 |
|   | Engine Speed (RPM) | 1200 | 1500 |
|   | Average Temperature at Intake Manifold (K) | 300 | 300 |
|   | Average Pressure at Intake Manifold (bar) | 2 | 2 |
| 6 | Diesel Equivalence Ratio (-) | 0.4 | 0.4 |
|   | Natural Gas Equivalence Ratio (-) | 0.3 | 0.5 |
|   | EGR fraction (-) | 0 | 0.5 |
|   | Reference CA50 (CAD) | 8 | 8 |

# Nomenclature

| Coefficient | Definition |
|---|---|
| $BD$ | Crank angle during burn duration |
| $CA50$ | Crank angle at 50% of fuel mass burnt |
| $CA50_{ref}$ | Reference CA50 |
| $EGR$ | Exhaust Gas Recirculation Fraction |
| $k_c$ | Polytropic constant |
| $m_{air}$ | Air mass entering the cylinder |

| Coefficient | Definition |
| --- | --- |
| $m_\text{diesel}$ | Mass of injected diesel |
| $m_\text{egr}$ | EGR mass entering the cylinder |
| $m_\text{NG}$ | Mass of injected natural gas |
| $m_\text{residual}$ | Mass of residual gas in the cylinder |
| $N$ | Engine Speed |
| $P$ | In-cylinder dynamic pressure |
| $P_\text{IVC}$ | Pressure at intake valve close |
| $P_\text{SOI}$ | Pressure at start of injection |
| $SOC$ | Crank angle at start of combustion |
| $SOI$ | Crank angle at start of fuel injection |
| $T$ | In-cylinder dynamic temperature |
| $T_\text{IVC}$ | Temperature at intake valve close |
| $T_\text{SOI}$ | Temperature at start of injection |
| $V$ | Dynamic volume of cylinder |
| $V_0$ | Cylinder volume at 0 crank angle degree |
| $V_\text{IVC}$ | Cylinder volume at intake valve close |
| $V_\text{SOI}$ | Cylinder volume at start of injection |
| $x_{O2,amb}$ | Oxygen mass fraction of the ambient air |
| $x_{O2,int}$ | Oxygen mass fraction at the intake manifold |
| $x_{O2,exh}$ | Oxygen mass fraction at the exhaust manifold |
| $x_b$ | Mass fraction of burnt fuel |
| $X_r$ | Mass fraction of residual gas |
| $X_d$ | Mass fraction of dilution |

| Coefficient | Definition |
|---|---|
| $\tau$ | Arrhenius function |
| $\phi_{DI}$ | Equivalence ratio of diesel |
| $\phi_{NG}$ | Equivalence ratio of natural gas |
| $\theta$ | Crank angle |

# Acronyms

| | |
|---|---|
| aTDC | After top dead center |
| BD | Burn duration |
| CA50 | Crank angle at 50% fuel mass burnt |
| CAD | Crank angle degree |
| ECU | Engine control unit |
| EGR | Exhaust gas recirculation |
| EVC | Exhaust valve close |
| EVO | Exhaust valve open |
| IMEP | Indicated mean effective pressure |
| IVC | Intake valve close |
| IVO | Intake valve open |
| KIM | Knock integral model |
| MKIM | Modified knock integral model |
| NG | Natural gas |
| PI | Proportion integral |
| PID | Proportion integral derivative |
| RMSE | Root mean squared error |

| | | |
|---|---|---|
| RPM | Revolutions per minute | |
| SOC | Start of combustion | |
| SOI | Start of injection | |
| st | Stoichiometric | |
| VGT | Variable geometry turbine | |

Tables

Table 1.      Engine Specifications

| | |
|---|---|
| Displacement Volume | 12.4L |
| Number of Cylinders | 6 |
| Compression Ratio | 17:1 |
| Valves per cylinder | 4 |
| Bore | 126mm |
| Stroke | 166mm |
| Connecting Rod length | 251mm |
| Diesel Fuel System | 2200 bar common rail |
| Air System | 2-stage turbocharger |
| IVC | -148.5 °aTDC |
| IVO | -363.5 °aTDC |
| EVO | 137 °aTDC |
| EVC | 389 °aTDC |

Table 2.      Range of Parameters in Simulation Data

| Quantity | Minimum Value | Maximum Value |
|---|---|---|
| Engine Speed (RPM) | 1200 | 1500 |
| $T_{IVC}$ (K) | 372.56 | 408.87 |
| $P_{IVC}$ (bar) | 2.85 | 4.37 |
| Diesel Equivalence Ratio (-) | 0.2 | 0.5 |
| Natural Gas Equivalence Ratio (-) | 0.2 | 0.7 |
| EGR (%) | 0 | 50 |
| SOI (° aTDC) | -20 | -10 |

EGR: exhaust gas recirculation; aTDC: after top dead center; SOI: fuel start of injection.

Table 3. Parameters of CA50 Prediction Model

| | |
|---|---|
| $c_1$ | $1.0504 \times 10^{-4}$ |
| $c_2$ | $1.4958 \times 10^{-4}$ |
| $c_3$ | 0.2284 |
| $c_4$ | -0.2604 |
| $c_5$ | 9591.9 |
| $c_6$ | -0.5962 |
| $c_8$ | 0.8292 |
| $c_9$ | 0.0522 |
| $c_{10}$ | -0.9682 |
| $c_{11}$ | 1.3359 |
| $k_c$ | 1.0546 |

Table 4. Error Response of CA50 Prediction

| Error Source | Error Value | Standard Deviation of CA50 Prediction Error | Maximum of CA50 Prediction Error |
|---|---|---|---|
| No Error | - | 0.5317 CAD | 2.1902 CAD |
| $P_{IVC}$ | +0.05 bar | 0.5199 CAD | 2.2831 CAD |
| $P_{IVC}$ | -0.05 bar | 0.5493 CAD | 2.2530 CAD |
| $T_{IVC}$ | +5K | 0.5140 CAD | 2.3048 CAD |
| $T_{IVC}$ | -5K | 0.5570 CAD | 2.3273 CAD |
| EGR | +5% | 0.5533 CAD | 2.3861 CAD |
| EGR | -5% | 0.5206 CAD | 2.4774 CAD |
| Equivalence Ratio of Diesel | +10% | 0.5328 CAD | 2.7210 CAD |
| Equivalence Ratio of Diesel | -10% | 0.5700 CAD | 2.6293 CAD |
| Equivalence Ratio of NG | +10% | 0.5362 CAD | 2.1428 CAD |
| Equivalence Ratio of NG | -10% | 0.5273 CAD | 2.2416 CAD |
| $X_r$ | +0.03 | 0.5341 CAD | 2.0857 CAD |
| $X_r$ | -0.03 | 0.5313 CAD | 2.2951 CAD |

Figures

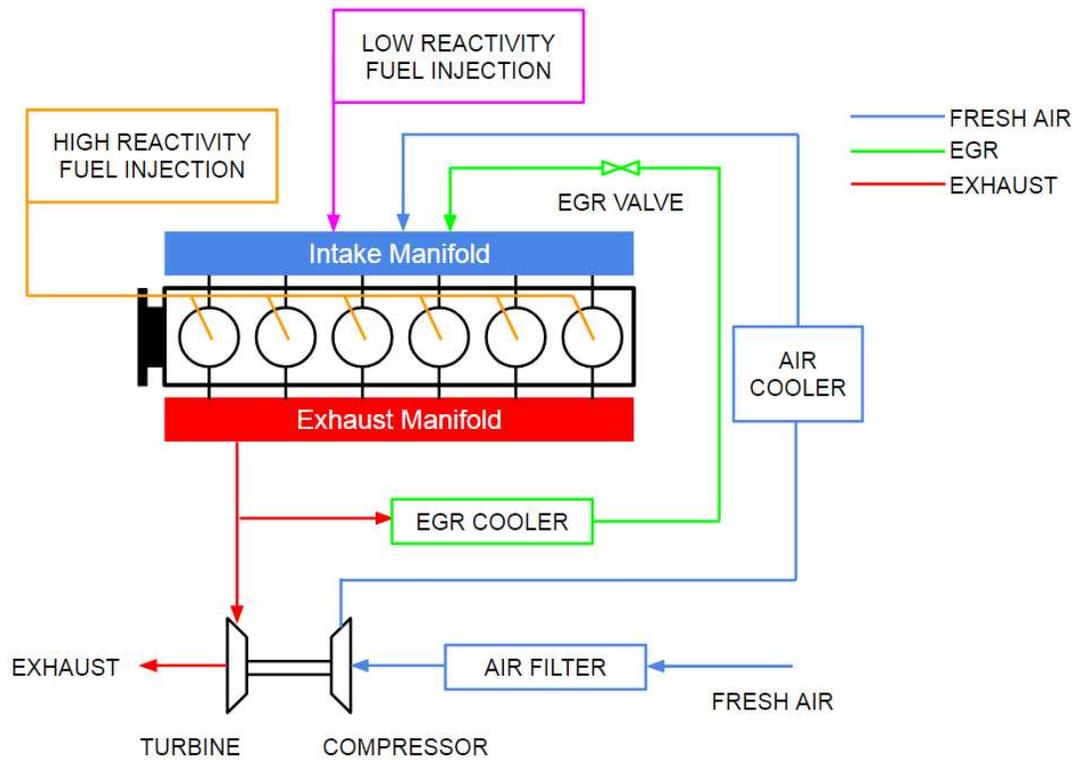

Figure 1. Schematic of dual fuel engine system

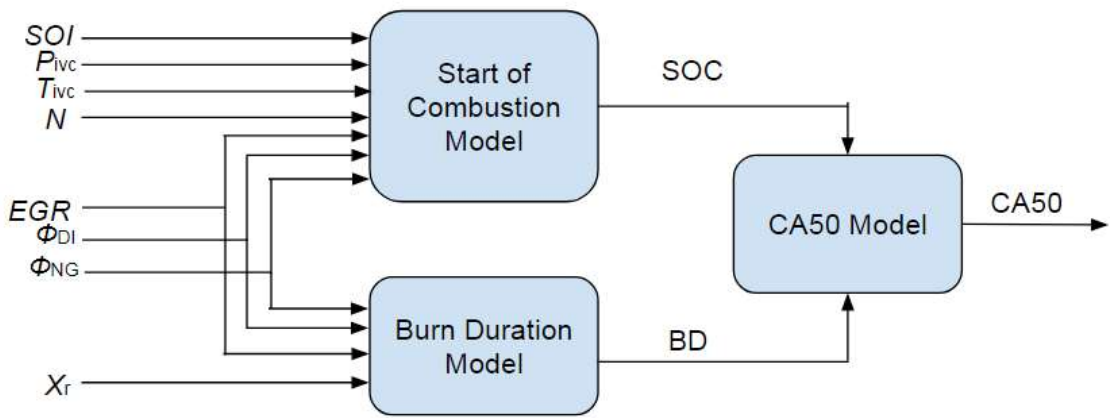

Figure 2. Block diagram of combustion phasing in dual-fuel engines

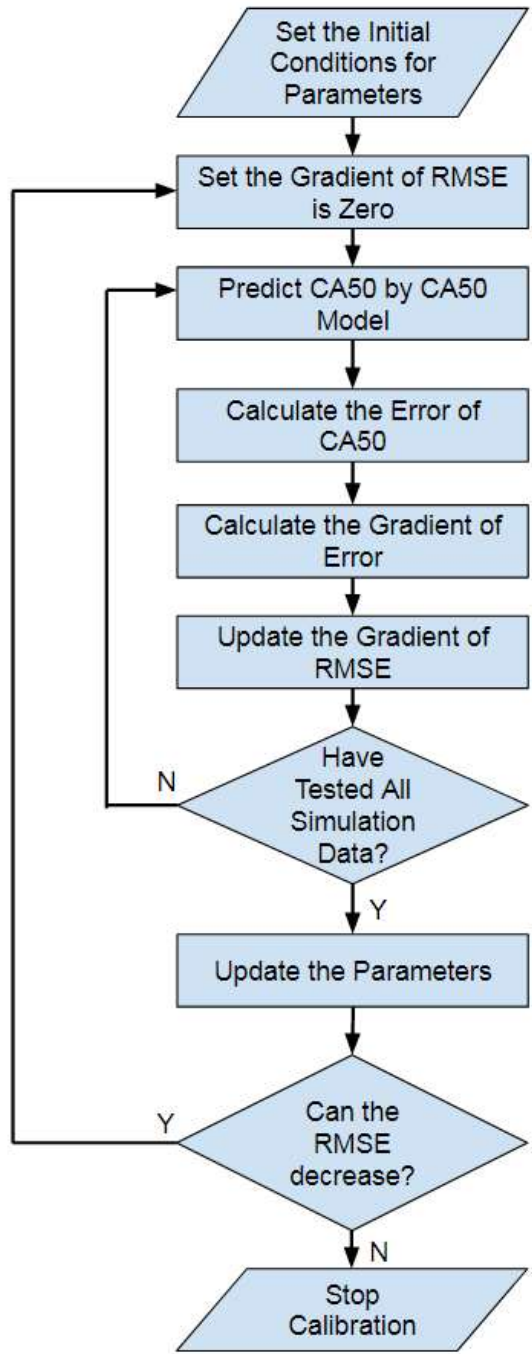

Figure 3.    Model calibration procedure

(a)

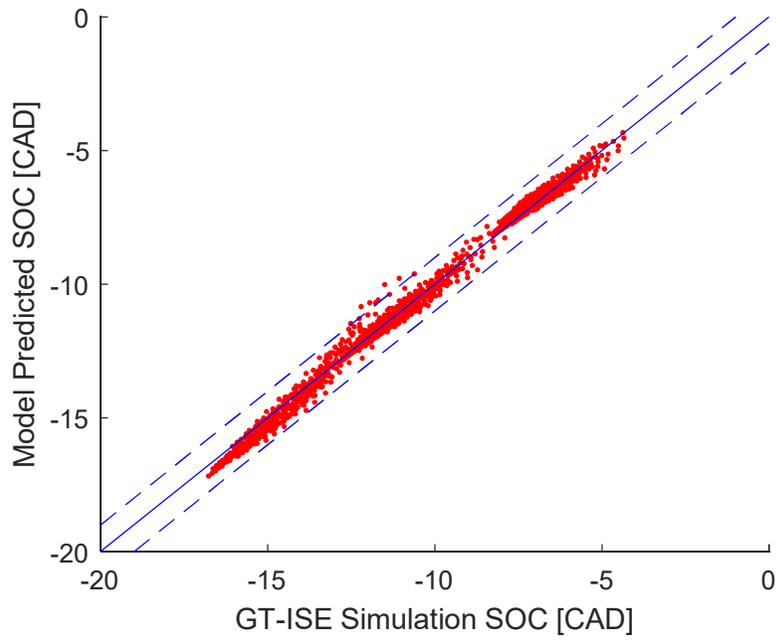

(b)

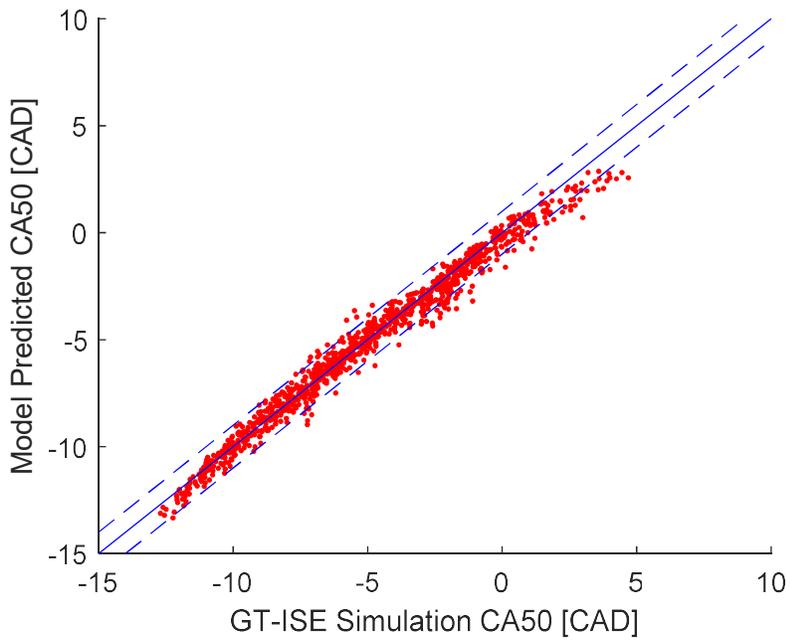

Figure 4. Comparison of model predicted (a) SOC and (b) CA50 with GT-ISE simulation

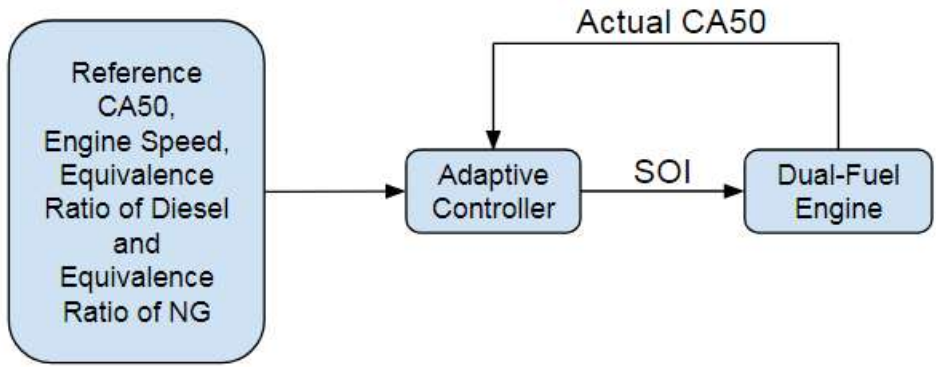

Figure 5. Block diagram of CA50 adaptive feedback control system

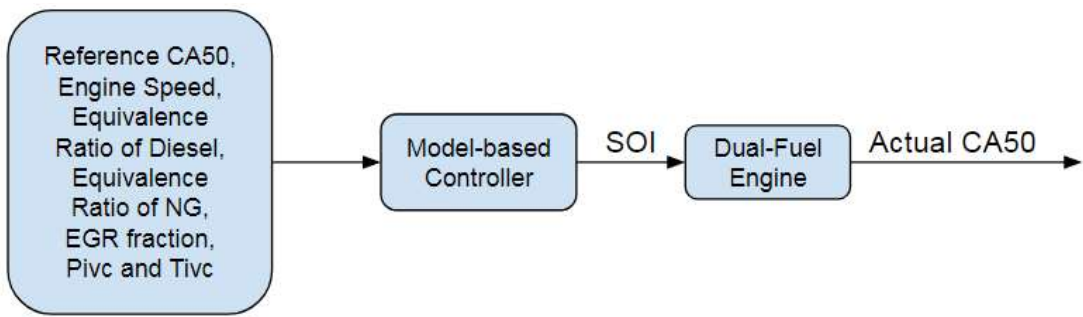

Figure 6. Block diagram of CA50 model-based open-loop control system

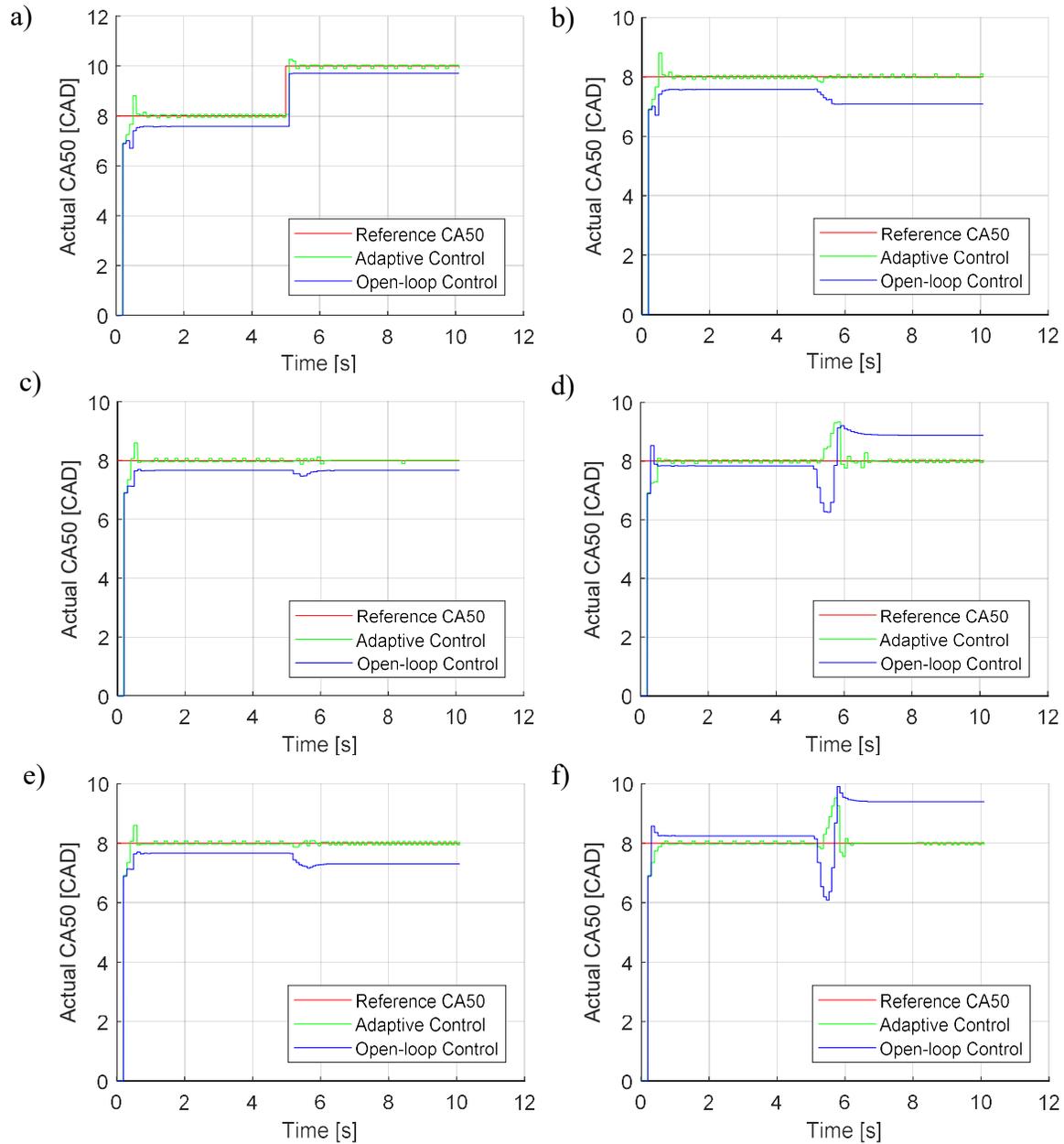

Figure 7. Controller performance for all simulation Cases: a) reference CA50 change, b) engine speed change, c) natural gas equivalence ratio change, d) EGR fraction change, e) combined natural gas equivalence ratio and engine speed change, and f) combined change in EGR fraction, engine speed, and natural gas equivalence ratio

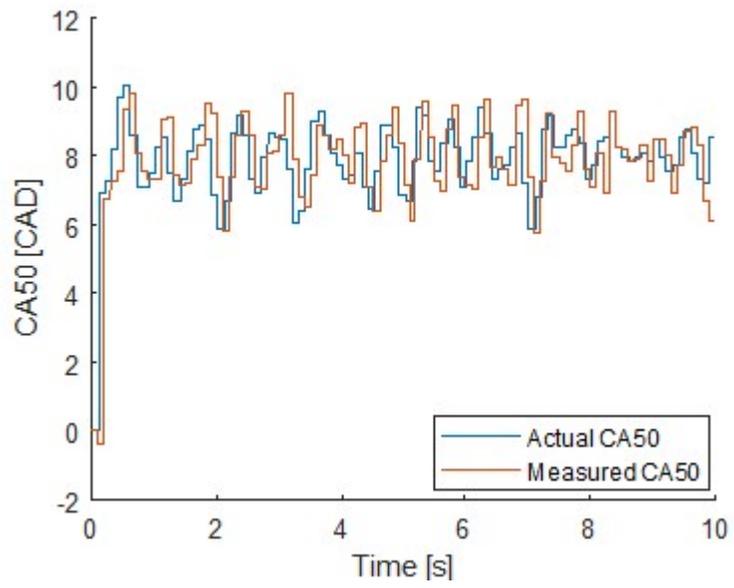

Figure 8. Simulation result of error response in adaptive controller